\documentstyle[prl,aps,twocolumn,epsfig]{revtex}

\begin{document}
\draft

\title{Quantum Anti--Zeno Effect}
\author{M. Lewenstein$^1$ and  K. Rz\c a\.zewski$^{1,2}$}
\address{$^1$Institut f\"ur Theoretische Physik, Universit\"at Hannover,
D-30167 Hannover, Germany}
\address{$^2$Center for Theoretical Physics and College of Science, 
Polska Akademia Nauk, 02-668
Warsaw, Poland}

\date{\today}

\maketitle

\begin{abstract}
We demonstrate that near threshold decay processes may be accelerated
by repeated measurements. Examples include near threshold photodetachment
of an electron from
a  negative ion, and spontaneous emission in a cavity close to 
the cutoff frequency, or in a photon band gap material.
\end{abstract}


\pacs{03.65.Bz}

The interplay between quantum dynamics and quantum measuments continues to
attract attention of physicists
 since the birth of quantum physics \cite{zurek}.
One of the classic examples of such an 
interplay is the, so called, quantum
 Zeno effect\cite{misra}. It consists in modifying  the quantum evolution
by the repeated measurments. The essence of the quantum Zeno effect is that 
repeated interrogation if the system is still in the inital state
 tends to quench the system in this state as the frequency of 
interrogations grows. 
The reason for that is that the quantum evolution generated by the 
hermitian  Hamiltonian is time reversible, and hence the probability
$p(t)$ of intial state 
occupancy behaves as $p(t)\simeq 1-(t/\tau)^2$  for short times $t$, 
with $\tau$ being 
the characteristic time scale. If we make $N$ interrogations within 
the time $T$, and if each of them gives  positive answer, i.e. collapses 
the actual state onto the initial state, then $p(T)\simeq (1-(T/N\tau)^2)^N \to
0$ as $N\to \infty$.
It is worth stressing, however, that for decay processes
into broad band reservoirs,
$\tau$ is typically of the order of the inverse of the reservoir width
(i.e. $10^{-16}$s for laser induced ionization of an atom).
For this reason, the Zeno effect in decay processes is very hard to observe
\cite{other}.

Following the idea of R. Cook \cite{cook}, the quantum Zeno effect has been
observed in a coherent process of  Rabi oscillations in a two level system
\cite{wine}. The idea of Zeno effect was also employed in recent
experiments on 
interaction free measurements and imaging \cite{kwiat}.  
The need for the collapse of the wave packet 
in the quantum Zeno effect has been critically analyzed in 
Ref. \cite{schenzle}. 

In this Letter we demonstrate that under suitable 
conditions, the decay of some  quantum systems from an initial state
might be accelerated
 by frequent interrogations whether  the system remains in that state.
We call this phenomenon a {\it quantum anti--Zeno effect}. The effect
takes place in systems that decay near  threshold. In fact, we show that even
below threshold, a frequent interruption of the evolution may enforce 
a complete decay of otherwise partially stable system. Somewhat 
similar effect has
been proposed by Gontis and Kaulakys\cite{litwa}, who argued that frequent
interrogations may destroy localization effect\cite{dima} in quantum 
chaotic systems, and may enforce classical stochastic motion.  

Near threshold decay occurs in several physical situations.
Here we discuss:
  a) near threshold photodetachment of an electron from 
 a negative ion \cite{nth},
b) spontaneous emission in a cavity close to cutoff frequency \cite{spont}, or
in a photon band gap medium\cite{band}. The dynamics  of all these systems
is governed by the following generic Hamiltonian
\begin{eqnarray}
 H&=&\hbar\Delta|0\rangle\langle 0|+\int \hbar\omega(k)
|k\rangle\langle k|\ dk \nonumber \\
&+&\hbar \int \left(g(k)|0\rangle\langle k|+ {\rm
h.c.}\right)dk,
\label{hamil}
\end{eqnarray}
where $|0\rangle$ is the decaying state (ground state for ionization,
excited state for spontaneous emission), and $k$ is a
 multi--index enumerating the states in the final continuum.
In the case a) of photodetachement these are just the
final states of an electron, in the case b) of spontaneous emission these
are products of the atomic ground state and single photon states 
characterized by a wave vector and a polarization denoted by $k$.
In Eq. (\ref{hamil}) rotating wave approximation (RWA) has been used
\cite{rwa}. 

In the
 case a), the detuning 
$\Delta$  is the difference between the ionizing laser
frequency  and the ionization potential divided by $\hbar$. 
The label $k$ can be identified with $\omega(k)\equiv \omega$, 
where $\hbar\omega$
is the energy of the photodetached electron. Various models of the coupling
$g(\omega)$ have been discussed in the literature. 
The angular momentum $l$
of the final states determines
the  behaviour of $|g(\omega)|^2\propto \omega^{l+1/2}$ for small $\omega$
in accordance 
with the Wigner's law\cite{wig}. In the following we will use the
expression 
\begin{equation}
|g(\omega)|^2= \frac{A}{\pi}\frac{\sqrt{\beta\omega}}{\omega + \beta},
\label{gk}
\end{equation}
for $\omega\ge 0$ in the  $s$---continuum. In the above expression
$\beta$ is the continuum width, and $A$, which is proportional
 to the laser intensity, is the Fermi Golden Rule photodetachment rate far
above the threshold\cite{nth}.

In the case b)  of spontaneous emission $\Delta$ is the difference 
between the transition frequency and the cavity cutoff, or band gap edge 
frequency, $\omega_c$. 
The $\omega(k)\ge 0$ describes the dispersion relation; near 
the boundary of the gap in the isotropic model it 
is well approximated by
\begin{equation}
\omega(k)= \frac{\omega_c}{k_0^2}(k-k_0)^2,
\label{disp}
\end{equation}
where $k_0$ is the wave vector at $\omega_c$.
In the relevant region of $k\simeq k_0$, $g(k)$ varies slowly, and 
can be taken as a constant proportional to the atomic transition dipole moment
$d$ \cite{spont,band}.

The survival amplitude in the initial 
state $|0\rangle$, $\alpha(t)$, maybe
written as an inverse Laplace transform
\begin{equation}
\alpha(t)=\int_{\Gamma}\frac{e^{zt}}{{\cal H}(z)}\frac{dz}{2\pi i},
\label{lapl}
\end{equation}
where  the countour $\Gamma$ runs parallely to the imaginary axis and is  
placed to the right from all singularities of the integrand, whereas
the resolvent function ${\cal H}(z)$ is given by
\begin{equation}
 {\cal H}=z+i\Delta + \int \frac{|g(k)|^2}{z+i\omega(k)}dk.
\label{rez}
\end{equation}

 In various regimes of parameters the following singularities contribute
to the time evolution.  Above the threshold, there exists a Wigner-Weisskopf
pole with a negative real part. The contribution of this 
pole represents exponentially damped oscillating term. Both above and below 
the threshold there is a contribution from the cut in the complex
plane.  The cut may be taken along the negative part of the real axis.
The cut contribution gives in particular
the celebrated algebraic long tail of the evolution
\cite{khal}. The threshold in the model does not correspond to $\Delta=0$,
and instead is dynamically shifted by 
\begin{equation}
\Delta_c= \int \frac{|g(k)|^2}{\omega(k)}dk. 
\end{equation}
For the model a) the threshold is shifted to 
 $\Delta_c=A$, whereas for the models b) it is pushed
toward infinity. The latter effect is caused by the fact that the
density of photon modes has a singularity for $\omega(k)=0$\cite{smooth}. 
Below the dynamically shifted threshold, there exists a purely imaginary pole, 
which contributes a non-decaying, purely oscillatory term to $\alpha(t)$.
Thus, sharp edge versions of the model b) exhibit always incomplete
damping. Of course, all of the above discussed contributions sum up 
to $p(t)=|\alpha(t)|^2\simeq 1-t^2/\tau^2$ 
for ultra short times.
It is the steep temporal behavior of the cut contribution for 
{\it moderately short times} which
we are going to exploit for the quantum anti--Zeno effect.
In fact, close to threshold the model a) enters the regime of asymptotic 
decay for such times, and decays generically as $1/t^3$. The exception 
occurs at the exact threshold, where the decay undergoes a
 crossover to the $1/t$ behavior and significant ``critical'' slowing down.
In particular, it is easy to see that if at such time  scales
$p(t)\simeq (\tau/t)^\nu$, for $t>\tau$ 
with $\nu>0$, then after $N$ interrogations 
within time $T$, the survival probability
$p(T)\simeq  (N\tau/T)^{N\nu}$ decreases
to values of the order of $\exp(-\nu e^{-1}T/\tau)$
as $N$ grows from 1 to $\simeq  e^{-1}T/\tau$.

The analytic expressions for $\alpha(t)$ can be easily obtained, and 
in the case a) and the model of Eq. (\ref{gk}) read
\begin{equation}
\alpha(t)=\sum_{i=1}^3\frac{y_i(y_i+\sqrt{\beta})}{3y_i^2+2\sqrt{\beta}y_i 
+\Delta}e^{(iy_i^2t)}{\rm erfc}[\exp(i\pi/4)\sqrt{t}y_i],
\label{exa1}
\end{equation}
where ${\rm erfc}(.)$ denotes the complementary error function of 
complex argument, 
and $y_i$ are the (complex in general) roots of 
\begin{equation}
y^3 + \sqrt{\beta}y^2 
+\Delta y+\sqrt{\beta}(\Delta-A)=0.
\label{pol1}
\end{equation} 
The corresponding result for the model b) is obtained by setting $\beta\to 0$,
$A\to\infty$ in such a way that $A\sqrt{\beta}\to \gamma^{3/2}$, where 
$\gamma = [\omega_c^{7/2}d^2/2\hbar c^3]^{2/3}$.

As in the Refs. \cite{cook,wine,groch}, we propose to interrogate the decaying
 system using
a series of short pulses of light resonant
 with a dipole allowed transition between the 
initial state and some other bound state of the atom. 
We consider the case when the pulse 
is intense enough to generate a  large number of  fluorescence photons,
and much longer than the characteristic time scale of the decay.
In such a case  sufficient
decoherence between the initial state and the continuum is incurred
to interrupt and reset the coherent evolution \cite{schenzle}. If $N$ 
interruptions are made within time $T$, the survival probability becomes 
$p(T)\simeq|\alpha(T/N)|^N$.   

In Fig. 1 we have plotted the survival probability after $T=100/A$
with various number of 
interruptions (indicated) 
as a function of detuning in the vicinity of the 
dynamically shifted threshold. The upper thick curve corresponds
 to the  fully coherent, uninterrupted dynamics.
For the model a) the following parameters
 were used: $A=10^4$Hz,  and $\beta/A=10^6$.  Observation time was taken
equal $T=10^{-2}$s. We consider interrogation pulses of the duration 
$10^{-6}$s, i.e. 
sufficient to produce a lot of fluorescence photons for a typical 
dipole transition. Such pulses are short enough to be considered
as instantenous on a time scale of the decay ($10^{-2}$); 
up to 1000 of such pulses well separated by $10^{-5}$s 
can be used. Both below and above threshold the more frequent  interruptions
cause faster decay. The reason for that is explained in Fig. 2, which presents
the
survival probabilty in a logarithmic scale
at the exact dynamical  threshold as a function of
time for the cases without, and with 200 interruptions with the time 
$T=10^5/A$. In this case the coherent decay is particularly slow and
occurs on the time scale of few $10^5/A$.   
Clearly, the coherent decay is non-exponential, and already for 
very short times can be well approximated by the algebraic $1/t$ dependence,  
which has very large derivative for such times. Measurements set
  the system
back to the initial state, and essentially press it to exploit this very
steepy
initial part of the time dependence. 	In effect,  the decay 
becomes much faster and  exponential in character, $p(nT/N)\simeq
(\tau N/T)^{n\nu}$. Apart from this acceleration of the decay, the 
measuments destroy the stability of the decaying state below threshold.
Even though,  in such a case the  
coherent evolution leads to non-vanishing
probability of survival as $t\to\infty$, however, 
 the frequent interruptions lead 
to the total decay. One should stress, the advantage of working with
the model a) is that the time scale can be here cotrolled, since
 $A$ is proportional to the ionizing laser intensity. Values between 
$10^4$ to $10^8$ are experimentally feasible. Being close to threshold 
requires laser stability within the A range (i.e. in the worst case 
in the kHz range). Observation times are reasonable, and 
``quantum jumps'' measuremt scheme could be implemented without
problems.
All that means, that quantum anti-Zeno effect
in near threshold decay is easier to observe experimentally than its
ultra short time analog -- the quantum Zeno effect.

In principle, the same effects as discussed for the model a) can be 
observed in the case b), but here the time scales are not so favorable.
depending whether we consider spontaneous decay in the cavity, or 
in the photon band gap medium, the reasonable values of $\gamma$ lie 
between $10^5$Hz (Rydberg atoms in a microwave cavity, or waveguide)
 and $10^9$Hz
(optical and infrared transitions in band gap media). In the model b) 
the characteristic time scale of the decay is typically $\simeq 10/\gamma$.
For Rydberg atoms the decay would take place on the scale of 
hundreds of microseconds, and could thus be monitored using 
``quantum jumps'' technique, which requires the time scales
 discussed above
in the context of model a). The decay of an impurity atom 
in the photon band gap medium  in the infrared or optical 
regime of photon frequencies would then take few nanoseconds, and would 
be rather hard to be interrupted by ``quantum jumps'' to another level.

It is worth stressing, that independently of the value of $\Delta$, 
the model b) leads to
a non-vanishing survival probability as $t\to \infty$. Moreover, this
final  survival probability depends very critically on the detuning,
and decreases very strongly below the threshold. This is illustrated
in Fig. 3 for $\Delta=-\gamma,0,\gamma$, which give final survival
probabilities $p_f\simeq$ .7, 0.45, and  0.15, respectively. Imagine that 
we perform ``quantum jump'' measurments in such a way, 
 so that the intervals between measurements are at 
least of the order of 100$\mu$s, i.e. larger than the time
scale of the evolution $\simeq 10/\gamma$. This condition is 
obviously fulfilled for an inpurity atom in the photon band gap
medium, and marginally fulfilled for Rydberg atoms in 
a cavity. On such a long time scale the evolution from Fig. 3 looks like
an instantenous initial slip follow by a constant.
 Depending on the detuning chosen 
the survival probability monitored by ``quantum  jumps'' will 
tend to zero rapidly with the number of interruptions $n$ as $p_f^n$.
{\it Frequent observing if the Schr\"odinger cat is alive, kills it faster}. 
Such experiment  requires ``tuning'' of
atomic transition frequency, or the photon band gap edge within the range of
$\gamma$, which can be for instance done using external static electromagnetic
field and Zeeman, or Stark effect.
 
Summarizing, we have shown that near threshold decay can be exploited 
to observe the quantum anti-Zeno effect. Typical near threshold 
decay processes are characterized by ultra  short period of 
quadratic decay, followed by a long phase of non-exponential decay, with a 
very large rate
 at the begining, gradually slowing down in the course of the dynamics. 
Monitoring if the decaying system remains in the initial state, shifts 
the system back to the fast initial phase of the non-exponential decay.
In effect, more frequent measurement, cause faster decay. We 
discussed experimental feasibility of 
observing this effect for two models describing a) near 
threshold  photodetachment 
of an electron from a negative ion, and spontaneous emission 
from  an atom 
located in a cavity, or from an impurity located within a 
photon band gap medium. Our scheme is quite general, and can be used
for any decaying system with several time scales. Frequent minitoring of such 
systems will tend to diminish the role of the long time
 time scales, and to blow up the role and influence of the
shorter time scales.
In particular, frequent monitoring  leads to  full decay of otherwise
partially stable systems. There are  implications of our result 
for quantum error correction schemes in quantum computing \cite{error}, which
typically employ quantum Zeno effect. Using computing units with non--standard
decay, such as the ones discussed above, might impose
 serious limitations on, or require developent of new error 
correcting strategies. 

The idea of this paper was born in fruitful discussions with Anna Sanpera.
We acknowledge the support of   Deutsche Forschungsgemeinschaft (SFB
407). K.R. thanks the Humboldt Foundation for its gereous support.
 
\begin{figure}[h]
\vspace*{0.8cm}
\caption{Survival probability for the model a) as a function of detuning after time $T=100/A$.
The thick upper curve corresponds to the coherent, uninterrupted evolution,
whereas the numbers of interruptions for other curves are indicated.
\label{survival}}
\end{figure}
 
\begin{figure}[h]
\vspace*{0.8cm}
\caption{Survival probability for the model a) 
as a function of time exactly at 
the dynamically shifted threshold.
The upper curve corresponds to the coherent, uninterrupted 
evolution,
whereas the lower  curves was obtained applying 50 (200) 
interruption pulses within the time $2\times10^5/A$, ($10^5/A$) respectively.
\label{cohe}}
\end{figure}

\begin{figure}[h]
\vspace*{0.8cm}
\caption{Survival probability for the model b) 
as the function of time. The values of $\Delta$ are indicated.
\label{modelb}}
\end{figure}


\begin{references}
\bibitem{rwa}To obtain
an accurate value of the  dynamical threshold shift $\Delta_c$ (see below), 
non-RWA corrections 
have to be considered \cite{nth}.
\bibitem{smooth} In realistic calculations, this singularity is always
 regularized, see for instance Ref. \cite{spont}.
\end{references}
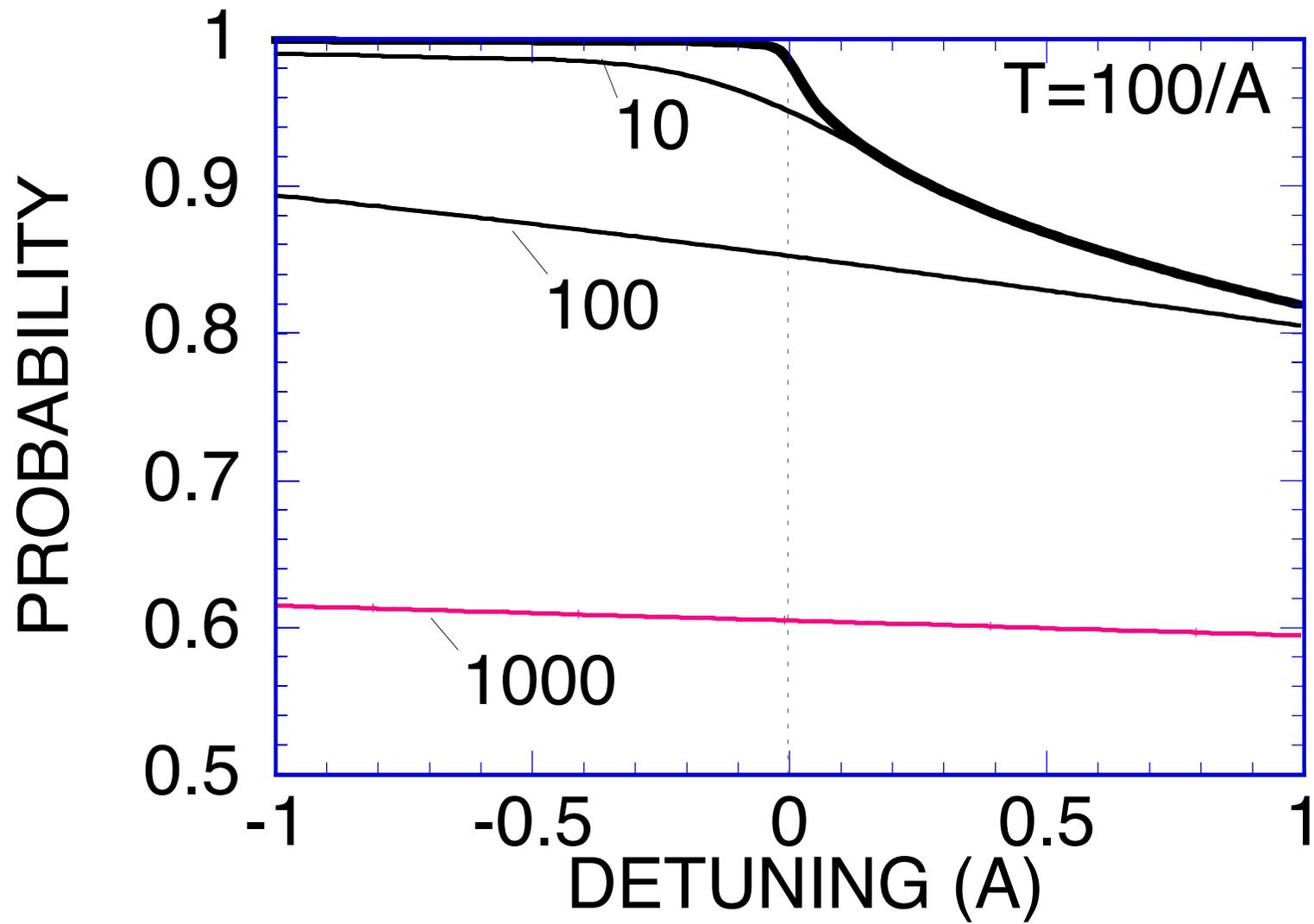
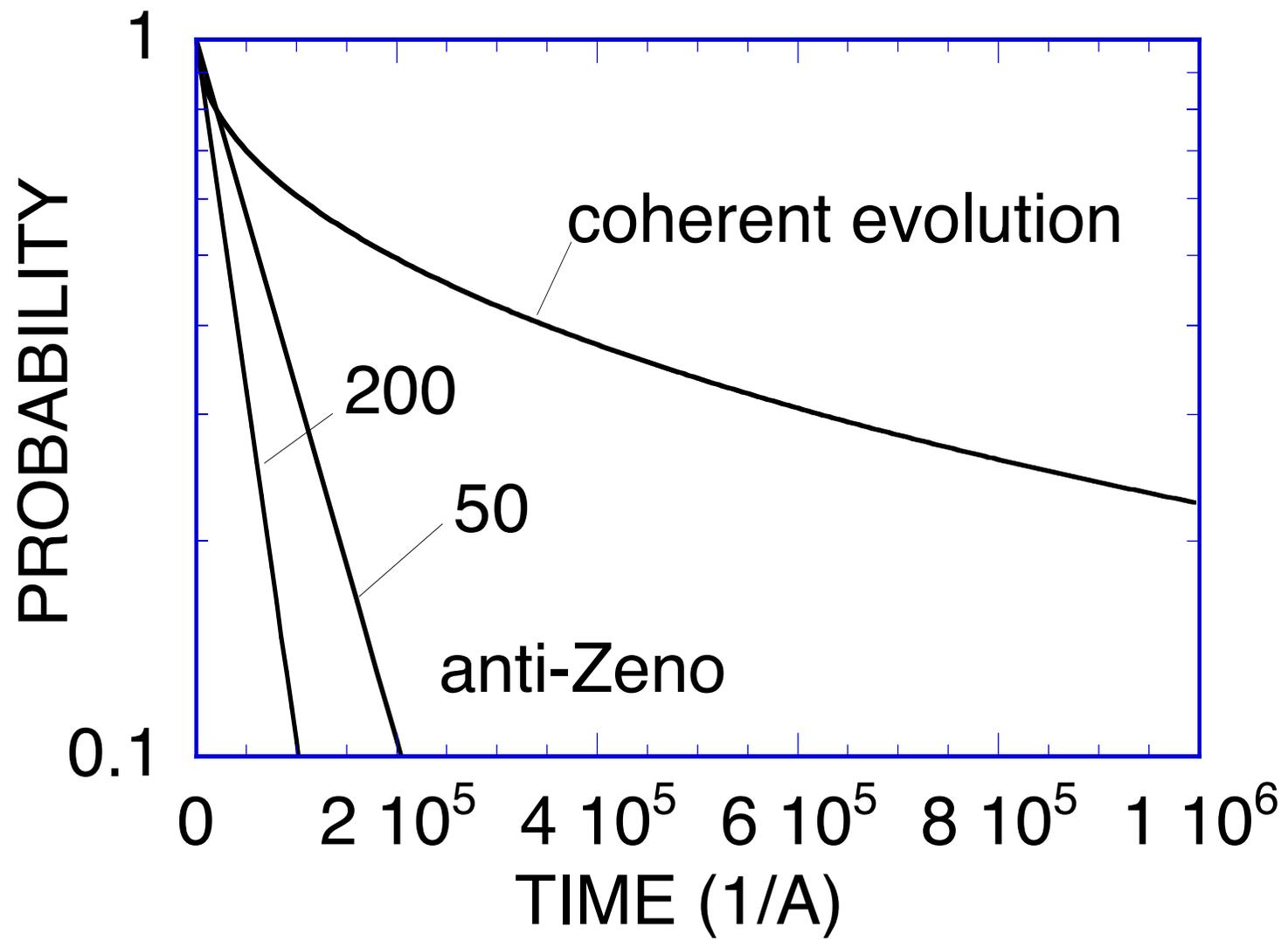
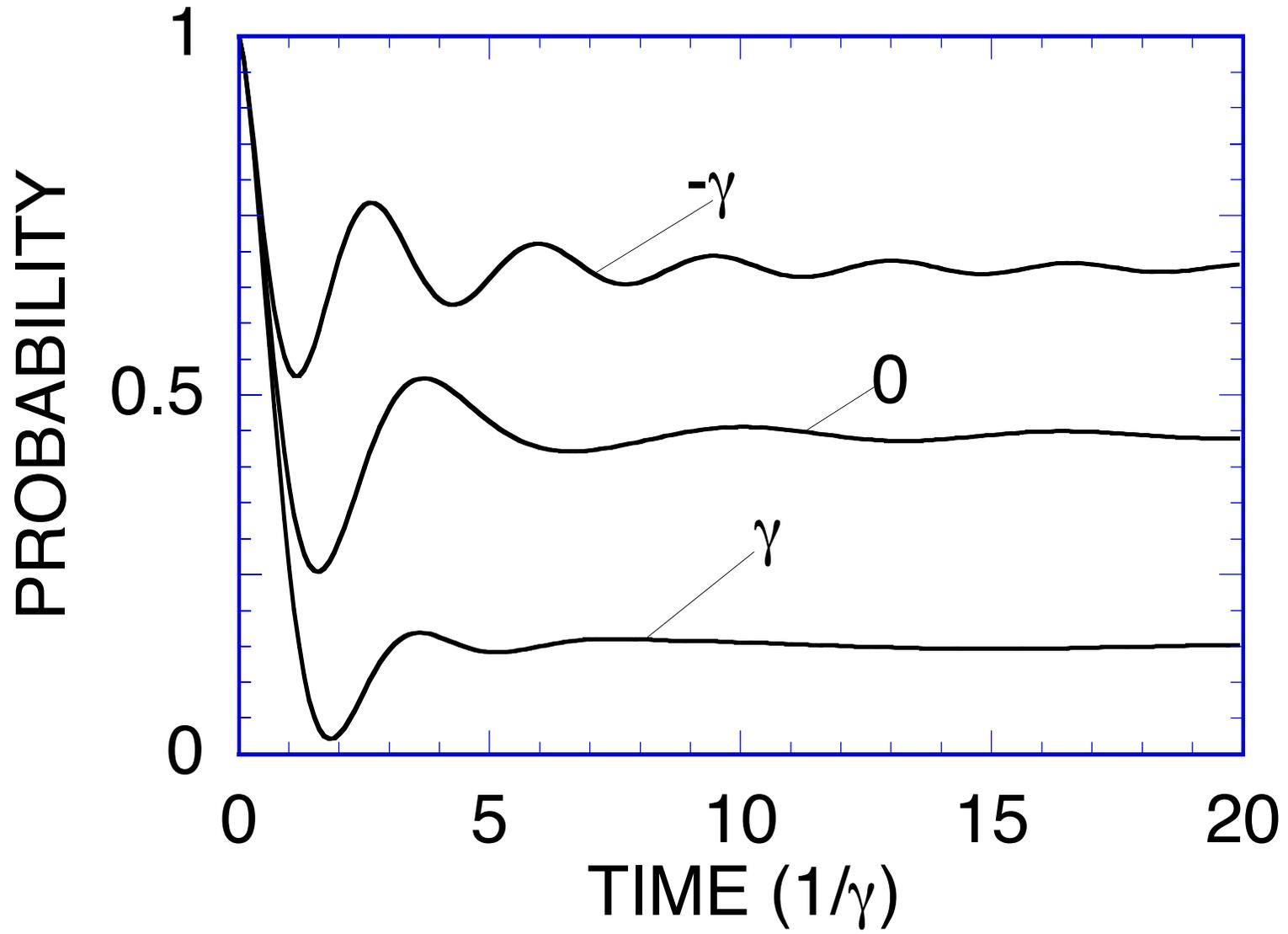
\end{document}